\documentclass[12pt,preprint]{aastex} 
\usepackage{graphicx,amssymb}
\usepackage{txfonts}
\usepackage{color}
\begin{document}
 \title{THE SOLAR ENERGETIC BALANCE REVISITED BY YOUNG SOLAR ANALOGS, HELIOSEISMOLOGY and NEUTRINOS}
   \author{Sylvaine Turck-Chi\`eze and Laurent Piau }

\shorttitle{The solar energetic balance}
\shortauthors{S. Turck-Chi\`eze, L. Piau and S. Couvidat}

 \affil{CEA/DSM/IRFU/Service d'Astrophysique, CE Saclay, 
              91191 Gif-sur-Yvette, France}

 \author{ S\'ebastien Couvidat   }
       
 \affil{  HEPL, Stanford University, Stanford, CA 94305, USA }
 \email{\it Sylvaine.Turck-Chieze@cea.fr; Laurent.Piau@cea.fr; Couvidat@stanford.edu}
 \date{Received December 21th 2010; accepted  February 25 th  2011}
 \begin{abstract}
The energetic balance of the Standard Solar Model (SSM) results from an equilibrium between nuclear energy production, energy transfer, and photospheric emission. In this letter, we derive an order of magnitude of several \% for the loss of energy in kinetic energy, magnetic energy, and X or UV radiation during the whole solar lifetime from the observations of the present Sun. We also estimate the mass loss from the observations of young solar analogs  which could reach up to 30 \% of the current mass. We deduce new models of the present Sun, their associated neutrino fluxes, and their internal sound-speed profile.
This approach  sheds  quantitative  lights on  the disagreement between the sound speed obtained by helioseismology and the sound speed derived from the SSM including the updated photospheric CNO abundances, based on recent observations. We conclude that  about 20\% of the present discrepancy could come from the incorrect description of the early phases of the Sun, its activity, its initial mass and mass-loss history. This study has obvious consequences on the solar system formation and the early evolution of the closest planets.
  \end{abstract}


 \maketitle
%
\newpage
\section{INTRODUCTION}
Impressive observational constraints unfolded during the past two decades to help check the physics of the solar interior. These efforts led to a remarkable agreement between two complementary probes: neutrinos and seismology \citep{Turck04a,Turck2010a,Turck2010d}. However, a reduction in the CNO solar photospheric composition has now been confirmed by two independent groups \citep{Asplund2009, Caffau08, Caffau09}  who directly extract it from photospheric and atmospheric lines, although a slight difference exists between them. This long-expected revision (see section 2.1 of  \cite{Turck1993}) produced an inconsistency between Standard Solar Model (SSM) predictions and the two aforementioned probes:  neutrino fluxes and radial sound-speed profile \citep{Turck04a,Bahcall05}. Several hypotheses were proposed \citep{Guzik, Basu08,Turck2008,Guzik2010}. Some of them  have now been rejected, like a large error in the new photospheric abundance estimate. Others are not currently easy to verify: 1) an incorrect microscopic diffusion: helium diffusion is, at least, measurable  but the extraction of radiation-zone CNO signatures in the equation of state casts some doubt on low subsurface values \citep{Lin07} even if this information is more difficult to obtain; 2) role of a fossil magnetic field on the bound-bound contributions and of the atmospheric magnetic field on the photospheric composition; 3) underestimated opacity calculations:  less than 10-15\% per element if elements other than CNO are incorrect, see \cite{Turck2010b}  ... The latter is stimulating extensive opacity comparisons and specific laboratory investigations \citep{Bailey, Turck2009}. Unfortunately such experiments are not yet available for the whole solar radiative zone (RZ).

In this letter, we examine another source of limitation of the SSM: the absence of solar activity effect during its evolution.  Our paper on the transport of momentum by rotation  showed the limits of the SSM to  predict the observed rotation profile \citep{Turck2010a} and the need for including magnetic field in the stellar equations \citep{Duez2010},  but we showed that their structural effects are small.  Strong activity phenomena are now observed on young solar analogs, and consequently a transformation of nuclear energy into kinetic and magnetic energy during the solar life must be studied. These phenomena produce a large variability in the X and UV spectra, inevitably accompanied by mass loss during the early phase of the solar evolution. The SSM does not take into account this dynamical aspect. Here we focus on the order of magnitude of such effects and on their past and present impacts. This letter is organized in three main sections. In section 2, we recall the SSM energetic balance and its limits. In section 3 we describe three updated solar models showing different types of loss of irradiance and mass loss connected to the past solar activity. The results are discussed in section 4.

\section{THE ENERGETIC BALANCE OF THE STANDARD SOLAR MODEL}
At the end of the eighties, the pp neutrino flux was easily estimated \citep{Spiro} and the results of GALLEX and SAGE \citep{anselmann92,gavrin92} could be interpreted as a first evidence that neutrinos might oscillate during their travel from the solar center to the Earth. Indeed, a reasonable estimate of the pp neutrino flux reaching Earth is: 
\begin{equation}
{\Phi \; \nu_{pp}} = 2 {L_\odot \over L\rm _{nucl}} *{1 \over {4 \pi d^2}}  
\end{equation}
where $L\rm _{nucl}$ is the energy produced by the pp chain, $L_\odot$ is the solar luminosity fixed at $3.846 \times 10^{33} \rm erg/s$ \citep{Guenther,Turck2010d}, and $d$ is the Sun-Earth distance. This equation derives directly from the fact that:
\begin{equation}
\rm 4p \;-> \;^4He + 2 e^+ + 2 \nu_e + 26.2 \,MeV  
\end{equation}
where 26.2 MeV is an approximate average value of the energy produced by the three chains ppI, ppII, and ppIII.
From these equations, the pp neutrino flux reaching Earth is $6.49\,10 \rm^{10}cm^{-2}/s$. This estimate is correct at the 10\% accuracy level, knowing that part of the solar neutrinos are also produced by the electronic capture of $^7Be$ and a small fraction ($2 \times 10^{-4}$) comes from the disintegration of $^8B$.  The detailed estimate of the different neutrino fluxes was improved in the framework of the SSM which allows to solve the four equations of stellar structure with extremely refined physics. The Sun is evolved up to its present age \citep{Guenther} of 4.60 $\pm$ 0.04 Gyrs, including 48 Myrs for the PMS. The solar luminosity increases by 30\% during this evolution. The transport of energy by photons, through interactions with the different species, is treated as properly as possible \citep{IglesiasRogers96}. Hydrostatic equilibrium is assumed during the evolution and no other source of energy is introduced. In the nineties, 
the predicted value of  $6 \times 10 \rm^{10}cm^{-2}/s$ at a 2\% accuracy level was obtained by different groups, see table 13 of \cite{Turck1993}.  This predicted value remained nearly unchanged these last years.
Therefore its only change results from the impact of the CNO cycle on the luminosity produced, which was slightly reduced by the recent CNO abundance update.

Then, SNO  definitely confirmed the existence of solar neutrino oscillations \citep{Ahmed}, measured a precise boron neutrino flux,  and consequently provided an accurate determination of the solar central temperature $T_c$. The inclusion of neutrino oscillations and the CNO abundance revision 
led the SSM pp, $^7Be$, and $^8B$ neutrino predictions to be only in marginal agreement with the neutrino detections \citep{Couvidat03,Turck2001a, Turck04a, Turck2010a}.  On the contrary, the evolved  Solar Seismic Model (SSeM), built to reproduce the observed sound speed, shows a remarkable agreement between its predictions and all the detection results (GALLEX-SAGE, CHLORINE, SNO, SUPER KAMIOKANDE and BOREXINO), but it produces central solar conditions different from the SSM (see \cite{Turck2010a} and Table 1).

More and more, seismic observations emphasize a dynamical view of the solar interior.  The Hale cycle of 22 years \citep{Hale38} seems to result from motions of the whole convective zone (CZ)
\citep{Brun04,Dikpati08}. In MHD simulations of this zone, the radial luminosity at a fractional radius r, can be expressed as the sum of different components:
\begin{equation}
{\rm L(r) \over {4 \pi r^2}} = \rm F_e + F_k + F_r + F_ u + F_v + F_m
\end{equation}
where $\rm F_e, F_k, F_r, F_ u, F_v, F_m$ are, respectively, the enthalpy, kinetic, radiative, unresolved eddy, viscosity, and Pointing, fluxes (see \citep{Brun04} for their definition).  These terms are ignored in the equations governing the SSM. If most of them are believed to be small in the CZ, none has been estimated in the RZ, in particular for the young Sun that could rotate 10-100 times faster than today, accompanied with related strong activity.

Unfortunately, 3D simulations cannot yet produce an evolutionary model of the global Sun. 
Nevertheless, it is now well established that  {\it the present constant luminosity}, name justified by the fact that in the SSM the solar luminosity varies by only 10$^{-8}$ during the last 100 years, is not confirmed by solar observations \citep{Frohlich06}. The adopted present luminosity is constrained at the $10^{-4}$ level in our models. The observed variability is directly correlated with the solar Hale dynamo which produces a mean $10^{-3}$ luminosity variation, with an additional 3-4 $\times 10^{-3}$ variation during the period of maximum activity of about 4-5 years due to the presence of preferred longitudinal spots appearing at the solar rotation periodicity. This variability can be accompanied by longer term variations  \citep{Turcklambert,Turcklefebvre}.

\section{A REVISED ENERGETIC BALANCE OF THE SOLAR INTERIOR}
The solar RZ encompasses $\approx 98\%$
of the total solar mass  and its microscopic ingredients (nuclear reaction rates,  opacity coefficients associated to the composition) imprint the sound-speed profile.  The
equilibrium between gravitational energy, nuclear energy production,
and the energy escaping by photon interaction, is achieved in the radiative interior
over long time scales. The sound-speed profile down to $0.06 ~R_\odot$ has been extracted from the Global Oscillations at Low
Frequency (GOLF) \citep{Gabriel} and the Michelson
Doppler Imager (MDI) \citep{Scherrer} instruments onboard SoHO
\citep{Turck2001a, Couvidat03}, and has recently been confirmed by BISON and MDI \citep{Basu2009}. In the RZ, a very slow meridional circulation could be in action \citep{Turck2010a} and a stable fossil magnetic field could still be present due to the very low diffusion coefficients considered. Both can affect the energetic balance, even though the introduction of these processes in the stellar equations show that their present structural effects are really small  \citep{Duez2010, Turck2010a}. However a quantitative estimate of these dynamical processes on the energetic balance requires to describe the early evolutionary stage, the evolution of the solar-core rotation, and the magnitude and topology of the fossil field.  The present study attempts to give an order of magnitude of these different effects. 
 \begin{figure}
 \vspace{-3cm}
\hspace{3pc}\includegraphics[width=35pc]{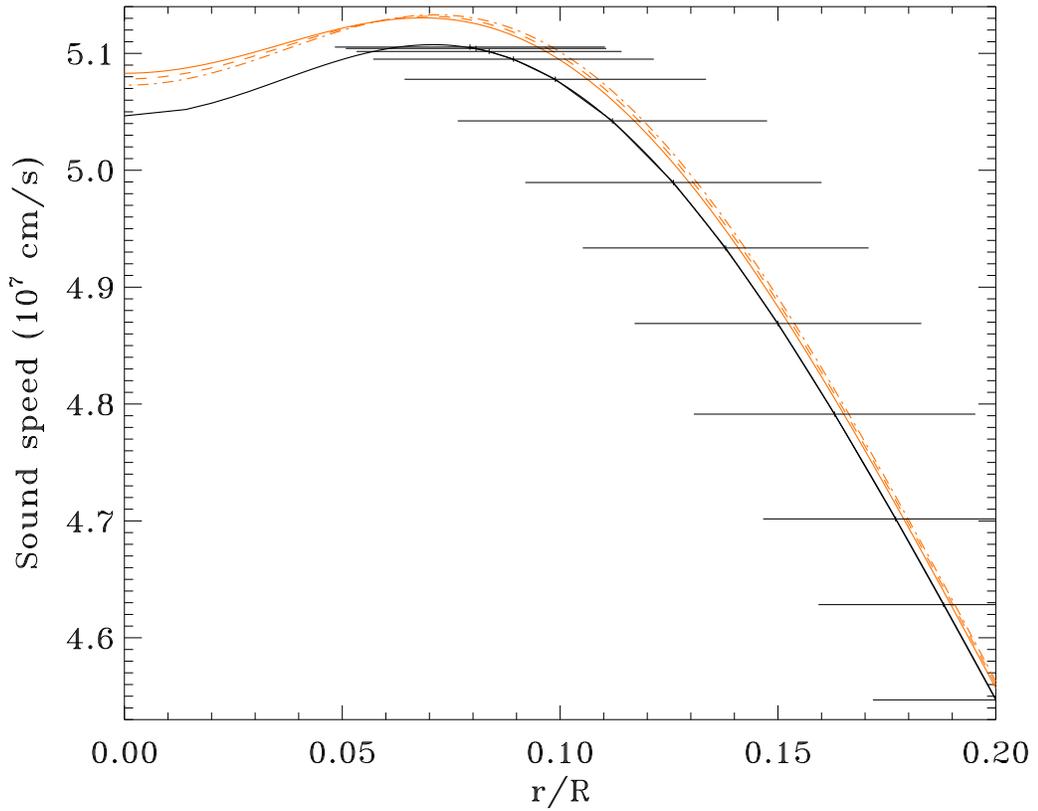}
\caption{  \small Sound-speed profile in the solar core corresponding to the seismic solar model (solid black line) in perfect agreement with helioseismic data, to the standard solar model with \cite{Asplund2009} composition (solid red line), to a model with luminosity increased by $2.5$\% (dashed red line), and to a model with luminosity increased by 5\% (dot-dashed red line). The vertical 3$\sigma$ error bars are very small and have been multiplied by 10 on this figure, while the horizontal error bars are rather large because few acoustic modes penetrate down to the core.\normalsize}
\label{fig:figure1}
\end{figure} 

First, we estimate the maximum energy loss  due to current radiative-zone dynamics. To this end, we use the central temperature difference between SSM and SSeM given in table 1. This central
temperature is now determined with an accuracy of $5 \times 10^{-3}$ thanks to the accuracy of neutrino fluxes. The pp luminosity varies like $T_C^4$, where $T_C$ is the central solar temperature. Therefore, an increase in $T_C$ of $1.5$\% could be interpreted as an increase in central luminosity of 6\% due to some energy loss by dynamical processes during the last million years, corresponding to the travel time of photons from solar center to the surface. This value is an upper limit because if the nuclear energy was partly transformed into another type of energy and evacuated, the Sun burnt more hydrogen and its density was also slightly increased.
Therefore we built two new solar models, for which we changed the calibrated luminosity at the present age by, respectively, $2.5$\% and 5\%. 
\begin{figure}
\hspace{3pc}\includegraphics[width=30pc]{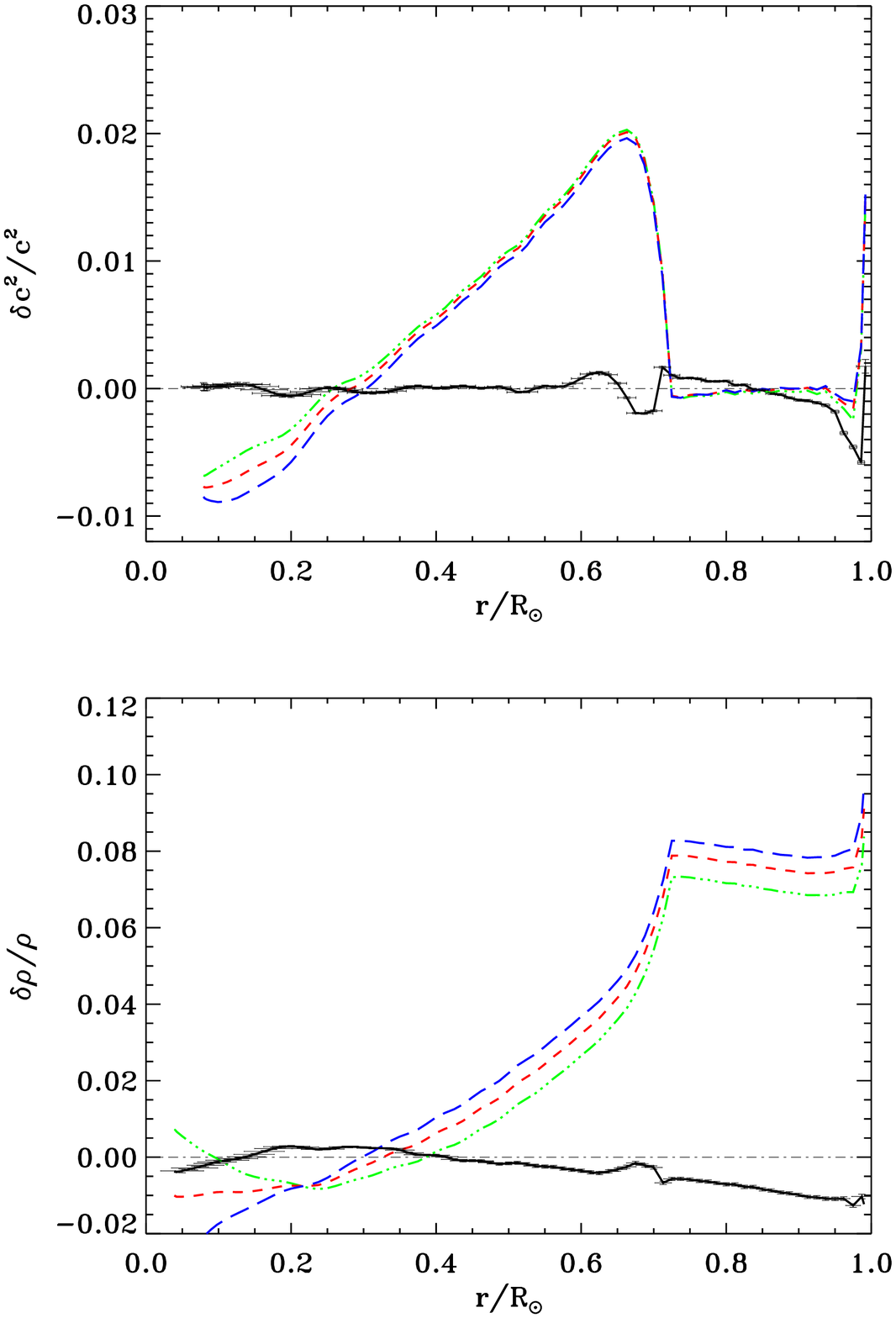}
\caption{  \small Squared sound-speed and density differences 
between the helioseismic inversions, obtained from the GOLF+MDI/SOHO acoustic modes, and various solar models: the standard solar model 
 with Asplund (2009) composition (green double dashed line), the 
seismic solar model (black line)  with the seismic error bars, and two models where a luminosity increase was introduced in the core (by 2.5 \% in red dashed line and by 5 \% in large dashed line). 
\normalsize}
\label{fig:figure2}
\end{figure}

The second estimate comes from the observation of young stars. In its infancy, the Sun was certainly more active and emitted more strongly in X and XUV. In the review of  \cite{Gudel}, the study of young solar analogs
shows how the evolved X luminosity varies with time with respect to the total bolometric luminosity. This wavelength range represents between $5 \times 10^{-7}$ and  10$^{-4}$ of the total solar luminosity $L_\odot$ and varies today 10-20 times more. Moreover \cite{Ribas} notes a variation of about 1000-2000 times the present emission in XUV for these young solar analogs. Therefore we applied the law described by these authors from their work on young stars:

\begin{equation}
 {dL (t) \over L} = \alpha  \tau (Gyrs)^{-1.23}
 \end{equation}
where $\alpha$ is equal to $1.31 \times 10^{-2}$, and $\tau$ is the age in Gyrs. This results in a luminosity loss of $\approx 3 \times 10^{-2}$
 at $0.5$ Gyrs which is similar to the previous estimate  when it is integrated over time because this loss decreases quickly with time.
 Thus the idea that the central solar region may be more evolved than what is today assumed in the classical stellar equations, is not such a stretch. 

Another consequence of solar activity is the associated mass loss. It is directly connected to the X luminosity.  In order to see the consequences of this effect, we used the law suggested by \cite{Wood2005} and we adopted the following decrease in mass for the first Gyrs: \begin{equation}
 \dot M _W= 1.0 \times 10^{-11} \tau (Gyrs)^{-2.23}  
 \end{equation}
 as recommended by \cite{Ribas}.
\begin{table}
  \begin{center} 
   \caption{Boron neutrino flux prediction (expressed in 10$\rm ^6 cm^{-2} s^{-1}$) compared to the measured SNO value of $5.21\pm0.45$, central 
   temperature $T_C$, initial helium content $Y_0$, central density $\rho_C$, central composition in hydrogen X$_C$ and helium Y$_C$, surface helium Y$_S$, and $\alpha$ parameter, for the seismic solar model, and the standard and non-standard solar models. All the SSM are computed with the Asplund (2009) composition.}
\vspace{5mm}    
     \begin{tabular}[h]{lccccc}
  \hline  
Model	& Boron  neutrino flux    &	$T_c$	& 		$Y_0$       &    $\rho_C$		\\

   \\
  SSeM     & 5.31 $\pm$ 0.6 & 15.74  & 0.277  &  153.02    &  \\
        & $X_C$= 0.339     &$Y_C$= 0. 645   &$\alpha_{MLT}$=  2.04 & $Y_S$= 0.251 \\
        \\
       
             SSM   &   4.50    &  15.54    &0.2645   & 150.6  \\
        &   $X_C$=   0.357 &  $Y_C$= 0.627  & $\alpha_{CGM}$ = 0.786  & $Y_S$= 0.235\\
 \\
SSM  L+0.025 &     5.32        &       15.67        &       0.2670        &        153.8         \\

                    & $X_C$=    0.347       &  $Y_C$=    0.637            & $\alpha_{CGM}$=    0.810         &$Y_S$= 0.237           \\

\\
              SSM  L+0.050 &   6.33    &  15.79  &         0.269          & 157.0   \\
            & $X_C$= 0.338    &  $Y_C$= 0.647     & $\alpha_{CGM}$= 0.836  &  $Y_S$= 0.239      \\
\\
 SSM  Tac &   4.2  &   15.49    &  0.2671   & 150.3 \\
      & $X_C$=  0.362   &   $Y_C$=  0.623  &$\alpha_{CGM}$= 0.712   &  $Y_S$=0.240 \\
\\          
SSM  M$_{init}$=1.33   &         4.74               & 15.58   &       0.2603    &    154.1   \\
         & $X_C$= 0.347   &  $Y_C$=0.637    & $\alpha_{CGM}$= 0.829       & $Y_S$=   0. 234    \\      
 \hline         
 \hline
 \end{tabular}
  \end{center}  
\label{tab:table1}
\end{table}

All the solar models of the present study include the recent inputs on reaction rates \citep{Turck04a}, the recent composition of  \cite{Asplund2009} and the equation of state and opacity tables built with this composition.   We also introduced the formalism suggested 
by \cite{Canuto1996} that improves the description of convection
(see references in \cite{Piau2011}).
\begin{figure}
\hspace{3pc}\includegraphics[width=30pc]{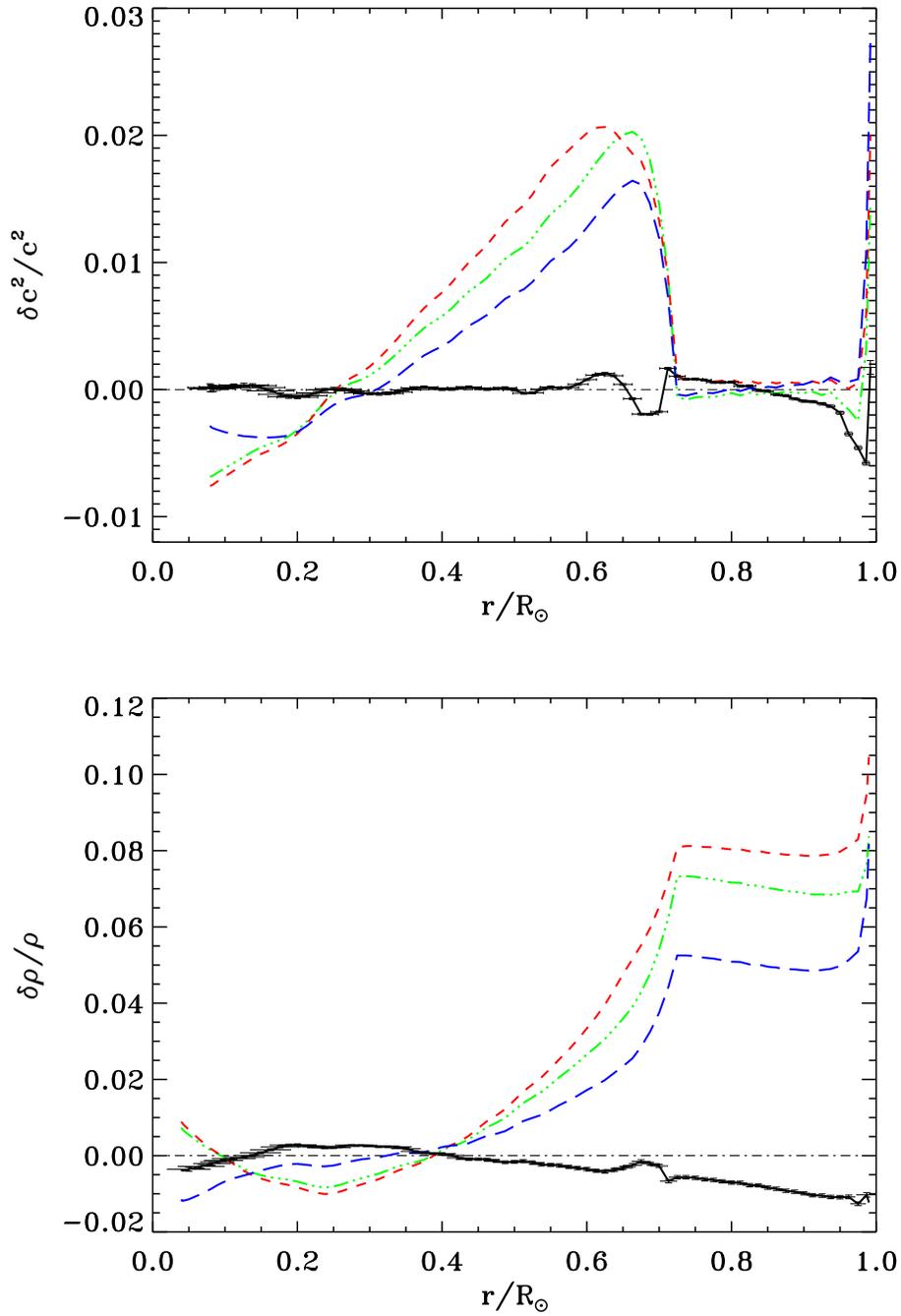}
\caption{  \small Same as Figure 2. 
The updated standard solar model 
is in green double dashed line, the SSM model with mixing in the tachocline is in red dashed line, the 
seismic solar model is in black line with the seismic error bars, and a model with mass loss in the early evolutionary phases following observations from young solar analogs is in blue long dashed line. \normalsize}
\label{fig:figure3}
\end{figure}
All the  results are summarized in Table 1 and Figures 1, 2, and 3.  

Table 1 lists the models computed. The  evolved SSeM, built to reproduce the observed sound speed, agrees with all the observables: neutrino fluxes and photospheric helium abundance. The new SSM shows a clear disagreement for the photospheric helium content. The discrepancy on helium is reduced by 20\% with the introduction of mixing in the tachocline following \cite{Brun2}. The SSM models with a central luminosity increased by, respectively, 2.5 and 5 \%  through a change in calibration show a net increase in the central temperature, and consequently an improvement of the boron neutrino predictions even  though the second case overestimates the neutrino flux. The mass-loss impact is less visible on the temperature than on the composition.

    Figure 1 shows the sound speed in the core obtained from GOLF \citep{Turck2001a} and confirmed by BiSON \citep{Basu2009}. While the vertical error bars are extremely small, the horizontal ones remain large in the region below 0.3 $R_\odot$ which contains more than half of the solar mass. This uncertainty  could be reduced by measuring mixed-mode frequencies of different degrees or high-frequency gravity modes \citep{Mathur}. By construction the seismic model (full black line) is in perfect agreement with the inverted sound speed from helioseismic data. We added three solar models: the SSM with turbulence in the tachocline  and the two models assuming a loss of luminosity of 2.5 and 5\% during the solar lifetime. These models (as shown in table 1) are in better agreement than the SSM in the actual core where the boron neutrinos are emitted. This figure shows the interest to better describe the solar core. More constraints in the inner part are needed not only to follow the dynamical processes but also to see if some dark-matter signature can be detected \citep{Lopes2010} which will differ from the signatures that we look for in the present study.
    
Figure 2  shows the squared sound-speed and density differences between helioseismic values and various solar models for the entire solar interior. The differences  for the SSM (dotted line) using the new photospheric composition and the  full spectrum of turbulence model of Canuto \mbox{et al.} (1996), remain important. In order to separate the effect of luminosity loss from the effect of mass loss, we show the models where the luminosity is increased by $+2.5$\% in small dashed  line and by +5\% in large dashed line. The effect is not large and has a sign opposite to the one in the core (see Figure 1). Using Equation 4 for the luminosity loss at the surface during the early stage, leads to a smaller effect because the star readjusts very quickly at each loss. The comparison of the two approaches shows that during the solar life an energy loss of a few \% is possible. However these simulations are too crude as the energy produced is not explicitly transformed into another energy type. Therefore Equation 4 alone cannot  reproduce the actual RZ.

In addition to SSM (in large dashed blue) and SSeM (in black), Figure 3 recalls the influence of the introduction  of mixing in the tachocline \citep{Brun2, Piau}. The horizontal turbulence in the tachocline slightly shifts the peak discrepancy downwards and suppresses the peak in density at the transition between radiation and convection. It reduces the discrepancy in the photospheric helium by 20$\%$ (see Table 1), but slightly decreases the central temperature. These points have already been mentioned in our previous works.
Figure 3 also shows the influence of mass loss given by Equation (5) on the present squared sound-speed and density differences.
 This mass loss is applied starting at 50 Myrs, corresponding to the beginning of the main sequence, and mimics the result of magnetic wind. The initial solar mass is 1.33 M$_\odot$ (where M$_\odot$ is the present solar mass). An even higher initial mass could be considered if the mass loss was applied at the separation of the solar disk near 3 Myrs. In the present study, the luminosity at the beginning of the main sequence is  $1.5$ L$_\odot$ instead of $0.7$ L$_\odot$, and the young Sun reaches 1 M$_\odot$ at $0.77$ Gyr. Its luminosity is smaller by 10\% than the present luminosity at $1.5$ Gyrs instead of 20\% with the SSM. In that case, the discrepancy on the squared sound-speed and density profiles is reduced and the predicted boron neutrino flux is slightly increased (table 1).
\normalsize

\section{SUMMARY, DISCUSSION  AND CONSEQUENCES}
This paper confirms the difficulty for the SSM to appropriately reproduce all the present solar observations: neutrino fluxes, sound-speed and density profiles, photospheric helium content, and location of the base of the convective zone.
This paper proposes two directions of investigation related to the activity of the young Sun  and to the comparison of central temperature between SSM and SSeM. We derive the order of magnitude of the corresponding effects and the quantitative implications on the aforementioned observables: (1) the Sun could have transformed about 2.5-4 \% of the energy produced  during the early evolutionary phases into another form of energy due to strong activity and the related interior dynamics, with a redistribution (or evacuation) of nuclear energy through kinetic and magnetic energies in the RZ; and (2) its initial mass could have been greater by 20-30\% than the present solar mass if the relation observed on young solar analogs is used.  Then, some residual effects on the present observables cannot be ruled out if we compare the difference of squared sound speed, the central temperature between SSM and SSeM and the boron neutrino fluxes. 

These ideas are not new \citep{Turck88,Sackman03,Guzik2010} but, for the first time, they are deduced from observations of young solar analogs and from the comparison of two solar probes, and are not only parametrized. Of course the mass loss studied here is rather large in comparison with previous works (limited by the previous agreement on the sound speed and lithium depletion) but is not ruled out  by observation of young solar analogs. The knowledge of the present rotation rate in the solar core could help put more constraints on the phenomena studied here.
This study will be pursued by a dedicated work on the history of the internal magnetism after introduction of the rotation history \citep{Turck2010a}. An active young Sun probably possesses a convective core and an early internal dynamo  in addition to some mass loss: this would greatly modify the initial conditions of planet formation and could partly solve the ``solar paradox'' mentioned a long time ago by \cite{Sagan1972} and never really totally solved: how such a low initial solar luminosity could be compatible with Mars early history  \citep{Forget97,Haberle98} ?  This young Sun and solar analogs would  also affect the poorly understood history of $^7$Li during their early evolution \citep{Piau}.  

Moreover, other ideas must be explored like the magnetic field effect on solar spectral lines which may impact the composition determination in the solar atmosphere \citep{Fabbian2010}, or the effect of a potential fossil field on opacities in the RZ, or the role of gravity waves.

\begin{acknowledgements}
 We thank the anonymous referee for his (her) suggestions and improvements of the first version.
\end{acknowledgements}

\bibliographystyle{apj}

\end{document}